\documentstyle[preprint,tighten,pra,aps]{revtex}
 \newcommand{\be}{\begin{equation}}
\newcommand{\ee}{\end{equation}}
\newcommand{\bea}{\begin{eqnarray}}
\newcommand{\eea}{\end{eqnarray}}

\newcommand{\ra}{\rightarrow}

\newcommand{\Lra}{\Longrightarrow}
\newcommand{\half}{\frac{1}{2}}
\newcommand{\Half}{\frac{3}{2}}

\begin{document} 

\pagestyle{empty} 

\title{\LARGE \bf Wigner-Eckart theorem in the crystal basis and
  the organisation of the genetic code}

\author{ A. Sciarrino}
\address{ Dipartimento di Scienze Fisiche,
Universit\`a di Napoli ``Federico II'' Ê\\
I.N.F.N., Sezione di Napoli  \\ 
 Complesso di Monte S. Angelo, Via Cintia, I-80126 Napoli, 
Italy  \\  e-mail: Sciarrino@na.infn.it}

\maketitle

\vspace{1cm}
\begin{center}
To appear in "Proceedings of VII Wigner Symposium" \\ 
University of Maryland, College  Park, 24-28 August 2001
\end{center}

\vspace{1cm} 

\begin{abstract}

Modelising the translation errors  by suitable mathematical operators 
in the crystal basis model  of the genetic code and requiring that codons 
prone to be misread  encode the same amino-acid,
the main features of the organisation in multiplets 
of the genetic code are described. 
 
\end{abstract}

\vfill
\vfill

\rightline{DSF-TH-34/2001}

\newpage  

\section{Introduction}
 
 The storage of genetic information is governed by the DNA, which is formed
 by four different nucleotides, characterized by their bases: 
adenine  (A) and guanine (G) deriving from purine, and cytosine (C) and 
thymine (T) coming from pyrimidine. In the double helix structure of DNA there is always a pairing
C-G and T-A. The transmission of information from DNA to build proteins
is a complex process of transcription and translation \cite{livre}. The flow of 
informations from DNA is transmitted  through the RNA  
  which also contains 4 bases, T being replaced by uracile (U).   
The mRNA ({\it messenger})  transmits the information from DNA to
the tRNA, which takes part to the protein synthesis. The transcription
from DNA to mRNA takes place following the bases complementarity rule:
$ A \ra U, T \ra A, G \ra C, C \ra G$ . In this context the idea of
{\it genetic code} emerges giving the connection law translating a 
sequence of nucleotides in the RNA (here and in the following we shall refer to
messenger) into a sequence of amino-acids (a.a.). A triple of nucleotides,
{\bf codon}, is read in the translation process. There  are
64 codons, which encode for the 20 a.a. (denoted in the following by their 
standard shortened notation), which are the building blocks for
any protein  and for the signal of the end of the protein synthesis (Stop
or Non-sense codons).  It follows that there is not an one to one correspondence 
between codons and a.a. and the code is degenerate exhibiting a complex
pattern of organisation in multiplets, ranging from  sextets to singlets,
in particular:  for the vertebrate mitochondrial code  (VMC) 
  2 sextets, 7 quartets and 12 doublets; for the 
eukaryotic or standard  universal, code (SUC) 
  3 sextets, 5 quartets, 2 triplets, 9 doublets and 2 
singlets  (seeTable(\ref{tablerep})). For a recent review with a rich bibliography 
to the original papers see \cite{Davies}. Codons encoding the same a.a. are 
called {\it  synonymous}. Since the discovery of the genetic code three, between
many others, puzzling questions have arisen: why  does  nature use  20  
a.a.   to build up proteins ? why has  the genetic code  
the peculiar  organisation in multiplets ? why is an a.a. encoded by
fewer codons than another ?  An answer to the last question
might be that  a.a. more frequently used are encoded by  greater 
multiplets. However such an explanation is weakly supported by
the analysis of the data, see Table \ref{tablefeq} see \cite{JTT}.  In particular 
the sextet encoding for Arg seems really overabundant. 

To explain the pattern of the genetic 
code at least six hypothesis have been put forward \cite{JH}:
\begin{enumerate}

\item the {\it frozen accident theory}, according to which the genetic code
is the result of a random event \cite{Crick}

\item the {\it stereochemical theory}, which first idea  dates back to 
1954, only one year after the discovery of the DNA by Crick and Watson,
and was suggested  by Gamow \cite{Gamow}, according to which the codons assignments 
 are the results of an affinity between the a.a. and the encoding codons
 \cite{Pelc}, \cite{Woese}
 
 \item the {\it coevolution theory}, according to which a.a. most
 closely related are encoded by close codons, i.e. codons differing by
 the change of one base \cite{Wong}
 
\item the {\it lethal mutation theory} according to which  
the genetic code  has evolved  minimizing the effect of point mutations
of the codons on proteins \cite{Son}, \cite{Epstein}

\item the {\it translation-error minimization theory} according to
which the evolution of the genetic code  has been governed by the minimization 
of the errors in the translation process \cite{Woese65}, \cite{GW}

\item the {\it genetic flexibility theory} according to which the genetic
code is the outcome of a balance between {\it robustness} and {\it
 mutability} \cite{MK}
 
\end{enumerate} 

However at my knowledege no quantitative model has been proposed to account
either for the number of a.a. either for the structure in multiplets.
Efforts have been concentrated to analyse the correspondence codons-a.a..
On this subject  a very large literature exists; it is now generally
accepted that this correspondence is not causal, but reflects the 
molecular structure of a.a., even  if there is a great  debate over the nature
of the dominant factors. However in almost all the papers on the subject the number
of a.a. and the structure in multiplets of the genetic code are assumed
as input  elements. The aim of this  talk is to propose a mathematical model
\cite{AS} able to reasonably explain the organisation of the genetic code, i.e. 
the number and the dimension of multiplets. In a way the proposed model
provides a mathematical frame for hypothesis 4) and 5).
 Clearly indeed a protection against the translation errors  is obtained if
 the codons more prone to be misread encode the same a.a..
 The first requirement to set  such a  mathematical model  
 is to identify codons as mathematical objects, this will be done
in the framework of the {\it crystal basis model} of the genetic code
\cite{FSS1},  which will be briefly recalled in the Sec. 3. To make the 
paper self-contained in Sec. 2 we recall the crystal basis for 
$U_{(q \to 0)}(sl(2))$. In  Sec. 4 the consequences of the model will
be  given without  details, which can be found in \cite{AS}.

\section{  Reminder of $U_{(q \to 0)}(sl(2))$}

For self consistence, let me recall the definition and main properties
of $U_{q \to 0}(sl(2))$ and of the $q$-tensor operator, see, e.g., 
\cite{BL}. ${\cal U}_q(sl(2))$  is defined by the following commutation relations
  
\be  
~[J_{+},\, J_{-}]  =  [2J_{3}]_{q}
\ee
\be  
~[J_3,\, J_{\pm}]  =  \pm \, J_{\pm} 
\ee
where 
\be 
~[x]_q = \frac{q^x - q^{-x}} {q - q^{-1}}  
\label{eq: 1}
\ee
In the following we shall omit the lower label $q$

The deformed enveloping algebra $U_{q}(sl(2))$ is endowed with an Hopf structure.
 In particular  the coproduct is defined by
\bea
\Delta (J_{3}) & =  & J_3 \otimes {\bf 1} + {\bf 1} \otimes J_3 
 \nonumber \\
\Delta (J_{\pm}) & = & J_{\pm} \otimes q^{J_3} + 
q^{-J_3} \otimes J_{\pm} \label{eq: 2}
\eea 
The Casimir operator can be written
 \be
 C = J_{+}  J_{-} \, + \, [J_{3}][J_{3} - 1] =
    J_{-}  J_{+} \, + \, [J_{3}][J_{3} + 1] 
 \ee 
  For $q$ generic, i.e. not a root of unity, the irreducible representations 
 (irrep.) are lalelled by
 an integer or half-integer number $j$ and the action of the generators
 on the vector basis $|jm>$, ($-j \leq m \leq j$) , of the IR is
\be
 J_{3} \,|jm> = m \, |jm>     \label{eq: 3}
 \ee
 \be
 J_{\pm} \, |jm> = \sqrt{[j \mp m] [j \pm m + 1]} \, |j,m \pm 1>
 = F^{\pm}(j,m) \,  |j,m \pm 1>  \label{eq: 4}    
 \ee    
 From eqs.(\ref{eq: 3})-(\ref{eq: 4}) it follows
 \be
 C \, |jm> = [j] [j + 1] \, |jm>
 \ee

 Let  us recall the definiton of $q$-tensor operator for  $U_{q}(sl(2))$. 
  An irreducible  $q$-tensor of rank $j$ is a family of $2j + 1$ 
operators $T_{m}^{j}$ ($ -j \leq m \leq j$) which tranform under
the action of the generators of  $U_{q}(sl(2))$  as  
\be
q^{J_{3}}(T_{m}^{j}) \equiv q^{J_{3}} \, T_{m}^{j} \, q^{-J_{3}} 
= q^{m} \, T_{m}^{j} \label{eq: C}
\ee
or 
\be
~[J_{3},\, T_{m}^{j} ]  = m \,  T_{m}^{j} 
\ee
\be
J_{\pm}( T_{m}^{j}) \equiv J_{\pm}Ê\, T_{m}^{j} \, q^{J_{3}} \, - \,
q^{-J_{3} \pm 1} \,  T_{m}^{j} \, J_{\pm} = F^{\pm}(j,m) \, T_{m \pm 1}^{j}
\label{eq: CO}
\ee
In deriving the above equations use has been made of the non trivial
coproduct eq.(\ref{eq: 2}).
The q-Wigner-Eckart ($q$-WE) theorem now reads:  
\be
<JM | T_{m}^{j} | j_{1} m_{1}> = (-1)^{2j} \, \frac{<J || T^{j} || j_{1}>}
{\sqrt{[2J + 1]}} \,<j_{1} m_{1} j m | JM> \label{eq: WE}
\ee 
   or
\be
 T_{m}^{j} \, |j_{1} m_{1}> = (-1)^{2j} \, \sum_{J = |j - j_{1}|}^{j + j_{1}} 
  \, \frac{<J || T^{j} || j_{1}>}{\sqrt{[2J + 1]}}  
  \,<j_{1} m_{1} jm | JM> \, |JM> \label{eq:WES}
\ee

Let us study now the limit $q \ra $.From the definition eq.(\ref{eq: 1})  we have
\be
  [x]_{q \ra 0}\, \sim \,  q^{-x + 1} \;\;\;\;\;\;\;\;\; \mbox{ $x \neq  0$}
 \label{eq: 5}
 \ee
 So it follows that
 \be
  F^{\pm}(j,m)_{q \ra 0} \, \sim \,  q^{-j + 1/2} \label{eq: 6}
  \ee 
   \be
  [j] [j + 1]_{q \ra 0} \, \sim \,  q^{-2j + 1} \label{eq: 7}
 \ee
  From eqs.(\ref{eq: 4}) and (\ref{eq: 6}) it follows that the action of the 
 generator $J_{\pm}$ is not defined in the limit $q \ra 0$.
  Let us define
 \be
 \widetilde{J}_{\pm}  = \Gamma_{0} \, J_{\pm} 
 \ee  
 where
 \be
  \Gamma_{0} = C^{-1/2} \label{eq: D}
 \ee  
  \be
 \Gamma_{0} \, |jm> =  ([j] [j + 1])^{-1/2} \, |jm>_{q \ra 0}  \sim \, 
 q^{j - 1/2} \, |jm>
 \ee
These operators are well behaved for ${q \ra  0}$. Their action
  in the limit $q \ra  0$ will define the {\bf crystal basis}: 
\bea 
\widetilde{J}_{+} \, |jm> & = & |j,m+1> \quad \mbox{for} \,\, -j \leq  m < j \\
\widetilde{J}_{-} \,  |jm> & = & |j,m-1>  \quad \mbox{for} \,\, -j < m \leq  j
\eea 
\be
\widetilde{J}_{+} \, |jj>  = \widetilde{J}_{-} \,  |j,-j>  =  0
\ee 
It is also possible to define a {\it Casimir} operator in the crystal basis
\be 
\widetilde{C}  = (J_{3})^{2} + \half \sum_{n \in {\bf Z_+}}
\sum_{k=0}^n (\widetilde{J}_{-})^{n-k} (\widetilde{J}_{+})^n 
(\widetilde{J}_{-})^k \,.
\ee
such that
\be
 \widetilde{C}  \, |jm> =  j(j + 1) \, |jm> 
 \ee
 Then I can define \cite{MS} $(q \to 0)$-tensor or crystal operator  
$\tau_{m}^{j}$ by :
\be
J_{3}(\tau_{m}^{j}) \equiv  
 m \, \tau_{m}^{j} \;\;\;\;\;\;
\widetilde{J}_{\pm} \, ( \tau_{m}^{j}) \equiv  \tau_{m \pm 1}^{j} \label{eq: QT}
\ee
Clearly, if $ |m| > j$ then $\tau_{m}^{j}$  has to be considered vanishing.
In the following I shall omit to explicitly write the tilde.  
The tensor product of two representations in the crystal basis
is given by \cite{Kashi}. 
{\bf Theorem} - If ${\cal B}_{1}$ and ${\cal B}_{2}$
are the crystal bases of the $M_{1}$ and $M_{2}$ ${\cal U}_{q
\rightarrow 0} (sl(2))$-modules, for $u \in {\cal B}_{1}$ and $v \in
{\cal B}_{2}$, we have:
\bea 
&& \tilde J_{-}(u \otimes v) = \left\{
\begin{array}{ll}
\tilde J_{-}u \otimes v & \exists \, n \ge 1 \mbox{ such that }
\tilde J_{-}^nu \ne 0 \mbox{ and } \tilde J_{+}^nv = 0 \\
u \otimes \tilde J_{-}v & \mbox{otherwise} \\
\end{array} \right. \\
&& \tilde J_{+}(u \otimes v) = \left\{
\begin{array}{ll}
u \otimes \tilde J_{+}v & \exists \, n \ge 1 \mbox{ such that }
\tilde J_{+}^nv \ne 0 \mbox{ and } \tilde J_{-}^nu = 0 \\
\tilde J_{+}u \otimes v & \mbox{otherwise} \\
\end{array} \right.
\eea 
So the tensor product of two crystal basis is a crystal basis and
the  states of the basis of the tensor space are {\it pure states}. In other words in 
the limit  $q \to 0$ all the $q$-Clebsch-Gordan  ($q$-CG) coefficients vanish 
except one which is equal to $\pm 1$. The Wigner-Eckart theorem 
eq.(\ref{eq:WES}) now reads
\be
 \tau_{m}^{j} \, |j_{1} m_{1}> = \, <J || T^{j} || j_{1}> \, |J,m_{1}+m> \label{eq: WE0}
\ee
where the value of $J$ ($ |j_{1} - j| \leq J \leq j_{1} + j$) depends on 
the value of $m_{1}$ and $m$ and of the order in which the tensor 
product of irreps. $(jm)$ and $(j_{1}m_{1})$. In the following the irrep. 
$(jm)$ will be considered as the second one. 

\section{The mathematical model}

 In the crystal basis model of the genetic code \cite{FSS1}  
 the 4 nucleotides  are assigned to the 4-dim   fundamental 
 irrep. $(1/2, 1/2)$ of $U_{q \to 0}(sl(2) \oplus 
sl(2))$ with the following assignment for the values of  the third 
component of  $\vec{J}$ for the two $sl(2)$ which in the following will be 
denoted as  $sl_{H}(2) $  and   $sl_{V}(2)$   :
\be 
	\mbox{C} \equiv (+\half,+\half) \qquad \mbox{T/U} \equiv 
(-\half,+\half) 
	\qquad \mbox{G} \equiv (+\half,-\half) \qquad \mbox{A} \equiv 
	(-\half,-\half) \label{eq:gc1}
	\ee
	and the codons, triple of nucleotides, to the $3$-fold tensor product of 
 $(1/2, 1/2)$.  We  report   in Table(\ref{tablerep}) the assignment of the 
 codons to the different irreps.
 The mathematical model {\it mimicking} the translation errors is 
 essentially based on the following  Assumption:
 
 { \bf Two codons are prone to translation error if their corresponding 
 states in the crystal basis model are connected by the action of a 
 suitable crystal tensor operators $\tau_{H,m}^{j} \otimes \tau^{j'}_{V,m'}$
 of $U_{q \to 0}(sl_{H}(2) \oplus sl_{V}(2))$ in the sense of the Wigner-Eckart
 theorem.}
 
 We assume, on phenomenogical grounds see \cite{Son}, \cite{FW}, \cite{FH}, that
  there is a
 hierarchy in the  occurrence of translation errors  and, in order of 
decreasing 
intensity, we consider:  
\begin{enumerate}
 \item  the transitions, in particular $C \ra U$ or $G \ra A$, concerning 
 nucleotides in the 3rd position 
  \item  the transversions, in particular  $C \ra G$, $U \ra A$ and $C \ra 
A$, in the nucleotides in 3rd position. 
  \item  the transitions (resp. transversions) concerning nucleotides 
in 1st position
  \item  the transitions (resp. transversions) concerning nucleotides 
in 2nd position
  \item  the  mutation induced by the transitions (resp. transversions) 
   on the first two nucleotides   
  \end{enumerate}
 Transitions (transversions) of the nucleotide in the middle position are 
 far weaker than transitions (transversions) in other positions.   
  
 The hierarchy in the  translation errors mechanisms means that
 a multiplet formed in a level is frozen; in the subsequent levels,
 the  merging of two  whole multiplets in a larger 
 structure is possible, if it is induced by the relevant tensor 
operator. If the transition is allowed only for some
 member of a multiplet, there is conflict between the choice of
 merging the multiplet in a larger one, so decreasing the variety
 of encoded a.a. but increasing the protection or preserving the
 multiplets decreasing the level of protection. In this case,  
 the formation of larger structures will generally take place or not 
according to the rule to protect the weakest codons, i.e. the codons more 
 inclined to be misread.  I assume that misreading of nucleotide C 
or A is the most common.  
Let me emphasize that we want to build the most simple model in which 
the codons, which are most subject to reading errors,  are synonymous; 
in this spirit the  explicitly analysed transitions ($C \ra U$, $G \ra A$) 
or transversions ($C \ra G$, $U \ra A$, $C \ra A$) have not to be 
considered as the  only possible misreadings, but as the representatives 
which allow the most simple modelisation.  
For simplicity  I consider only the transversions decreasing or leaving unchanged 
the value of $J_{H,3}$.  
 Finally, it is clear from the Table(\ref{tablerep}) that there are generally more than one 
irrep. labelled by the same value of $(J_{H}, J_{V})$ whose content in the 
constituent nucleotides is different. The transformation properties of a 
crystal tensor operator determine which states are related each other, 
only according to the irreps. to which the  states belong to.  To take
into account someway the multiplicity of irreps.
, generally, the choice of the  rank of the tensor 
operator will depend on the  position of the misread nucleotide and on
the  irrep. to which the codon belongs to.  
  The transitions and the  transversions are modelised by the following 
crystal tensor operator, the value of the component being determined by
the labels of the nucleotides, see eq.(\ref{eq:gc1}):
 \be
 C \ra U  \;\;\;\;\; \mbox{or} \;\;\;\;\;  G \ra A   
\;\;\;\;\;\;\;\;\;\; 
\tau_{H,-1}^{1}  \otimes \tau^{a}_{V,0} \label{eq:t1}
 \ee
 \be
 C \ra G \;\;\;\;\; \mbox{or} \;\;\;\;\; U \ra A   
\;\;\;\;\;\;\;\;\;\; 
\tau_{H,0}^{b}  \otimes \tau_{V,-1}^{1}  \label{eq:tr1}
 \ee 
\be
 C \ra A     \;\;\;\;\;\;\;\;\;\;  \tau_{H,-1}^{c} \, \otimes \, 
\tau_{V,-1}^{d} \label{eq:tr2}
 \ee 
where the values of the rank $a$, $b$, $c$  and $d$ depend on the position 
inside the codons  of the  misread nucleotide and on the irreps. to which 
the codons belong, see next section. 
The above choice  for the horizontal (resp. vertical) part of the 
crystal vector operator in eq.(\ref{eq:t1}) (reps. eq.(\ref{eq:tr1})) is 
indeed the most simple choice according to the change in the labels of the 
states of  
codons for transitions (resp. transversions), see eq.(\ref{eq:gc1}).
 The choice of the rank of the 
vertical (resp. horizontal) part of the crystal operator in 
eq.(\ref{eq:t1})  (resp. 
eq.(\ref{eq:tr1})), as well as the tensor operator modelising
the tranvsersion C $ \ra $ A,  is somewhat arbitrary.  
The value of the rank of the operator modelising the translational errors 
in 2nd position will be  generally assumed  larger than the one describing 
errors in 1st position and the latter one will be generally assumed larger 
than the one describing errors in 3rd position, so to model the less 
frequent misreading.

\section{ Outcome of the modelisation of translation errors}

In the following I use the 
standard notation: X,Y,N denoting any nucleotide, Y = C, U (pyrimidine),
R = G, A (purine).
\begin{enumerate}

\item{\bf Misreading of 3rd  nucleotide}

The  transitions in 3rd position in the codons XZC and XZG, 
  are modelised by the operator given by eq.(\ref{eq:t1})  with $a = 0$:  
(in the following the equations have to be read by the western 
rule from left to right) 
\bea
\psi(XZC)  \circ (\tau_{H,-1}^{1} \otimes \tau^{0}_{V,0})  \; & \Lra  
& \; 
\psi(XZU)   \label{eq:3t1}  \\ 
\psi(XZG)  \circ ( \tau_{H,-1}^{1} \otimes \tau^{0}_{V,0}) \; & \Lra  
& \; 
\psi(XZA)  \label{eq:3t2}
\eea
   We get  the 
splitting of the 64 codons  in 32 doublets of the form XZR and XZY.
 
The  transversions in 3rd position in the codons XZC and XZU 
 are modelised by the following operators:  
\bea
 \psi(XZC)  \circ (\tau_{H,0}^{b}  \otimes \tau_{V,-1}^{1})  \; & 
\Lra  & 
\; \psi(XZG) \label{eq:tva}  \\ 
\psi(XZU)  \circ (\tau_{H,0}^{b - 1}  \otimes \tau_{V,-1}^{1})  \; & 
\Lra  & 
\; \psi(XZA)  \label{eq:tvb} \\ 
 \psi(XZC)  \circ (\tau_{H,-1}^{b}  \otimes \tau_{V,-1}^{1} ) \; & 
\Lra  & 
\; \psi(XZA) \label{eq:tvc}
\eea  
where in eqs.(\ref{eq:tva}),(\ref{eq:tvb}),(\ref{eq:tvc})
 $b = 2$   if  the first two nucleotides (dinucleotide) XZ are: CA, GA, CG,
 UG, UA, UU, AU, AA,  GG, AG) t   and $b = 1$   otherwise.
  
We  get  the  merging of 16 doublets in 8 quartets,
 the  quartets being the codons whose the first two 
 nucleotides are: CC, CU, CG, UC, GG, GC, GU, and AC.
  
\item {\bf Misreading   of 1st  nucleotide}

The transitions  are modelised by the operators 
  eq.(\ref{eq:t1}) with $a = 1$:    
\bea
\psi(CXN)  \circ (\tau_{H,-1}^{1} \otimes \tau^{1}_{V,0})  \; & \Lra  
& \; 
\psi(UXN)  \nonumber \\
\psi(GXN)  \circ (\tau_{H,-1}^{1} \otimes \tau^{1}_{V,0})  \; & \Lra  
& \; 
\psi(AXN) \label{eq:to1}
\eea
    We get the merging of the doublet UUR and the quartet CUN in a 
sextet (encoding Leu).   

The transversions in first position are modelised by the operators:
\bea
\psi(CXZ)  \circ (\tau_{H,0}^{1} \otimes \tau_{V,-1}^{1})  \; & \Lra  
& \; 
\psi(GXZ) \label{eq:1tv1}   \\ 
\psi(UXZ)  \circ (\tau_{H,0}^{2} \otimes \tau_{V,-1}^{1})  \; & \Lra  
& \; 
\psi(AXZ) \label{eq:1tv2}  \\    
\psi(CXZ) \circ (\tau_{H,-1}^{c} \otimes \tau_{V,-1}^{1} ) \; & \Lra  
& \; 
\psi(AXZ)   \label{eq:1tv3}
\eea
where $c = 1$ if the codons CXZ and UXZ  belong to the same irrep.
  and $c = 2$ otherwise.   
  As a consequence the doublet AGR  merges into the quartet CGN forming 
  another sextet (encoding Arg).
  
 \item {\bf Misreading   of  central nucleotide}
   
 The transitions  are modelised by the operators 
  eq.(\ref{eq:t1}) with $a = 2$:   
\bea
\psi(XCN)  \circ (\tau_{H,-1}^{1} \otimes \tau^{2}_{V,0})  \; & \Lra  
& \; 
\psi(XUN)  \nonumber  \\
\psi(XGN)  \circ (\tau_{H,-1}^{1} \otimes \tau^{2}_{V,0})  \; & \Lra  
& \; 
\psi(XAN) \label{eq:to2}
\eea
 No modification of the previous established pattern comes out.  
     
       We modelise the transversions  as
\bea
\psi(XCZ)  \circ (\tau_{H,0}^{1} \otimes \tau_{V,-1}^{2})  \; & \Lra  
& \; 
\psi(XGZ)    \label{eq:2tv1} \\
\psi(XUZ)  \circ (\tau_{H,0}^{2} \otimes \tau_{V,-1}^{2})  \; & \Lra  
& \; 
\psi(XAZ)    \label{eq:2tv2} \\ 
\psi(XCZ) \circ (\tau_{H,-1}^{c} \otimes \tau_{V,-1}^{2} ) \; & \Lra  
& \; 
\psi(XAZ)   \label{eq:2tv3}
 \eea
where $c = 1$ if the codons XCZ and XUZ belong to the same   
irrep.  and $c = 2$ otherwise.
 It turns out that one should expect the fusion in a sextet of the 
quartet UCN and of the doublet UGY. This sextet does not appear, but as we shall 
see  below   the quartet UCN indeed merges with the 
 doublet AGR. One should also expect the fusion in a sextet of the quartet 
CCN and the doublet CAY, which indeed does not happen. 
 Both these results suggest that indeed the misreading of the central nucleotide
is a very weak effect, if not enhanced by the simultaneous misreading of 
the first nucleotide. 

\item {\bf Misreading  of two nucleotides}
  
   The transition  and transversion   of the first
   (second)  nucleotide  is modelised by the same operator used  
    for the translation or transversion on the first nucleotide.
     In the following we denote with a lower label the position of 
   the nucleotide where the operator acts.
   The action of the  two-nucleotides operators  has to be computed in
    the following way: as first step one has to compute the action of
 the operator labelled by I giving rise to a "virtual" state with the
    labels assigned by the action of the relevant operator on the 
    initial state of the codon, then one considers the action of the 
    operator labelled by II on the "virtual" state and gets the  labels
    of the final state. If these labels   denote, see Table(\ref{tablerep}),  
    the state corresponding to the codon, the  misreading is allowed.
    As an example let us compute
    \be
     \psi(CCN)    \;  \Lra  \; \psi(UUN) 
      \ee  
     
     \bea
     & \mbox{(i)} \;\;\;\; \psi(CCN) \circ 
    (\tau_{H,-1}^{1} \otimes \tau_{V,0}^{1})_{I}  \Lra  \psi(UCN)_{(vir)}
    \nonumber \\
    & \mbox{(ii)} \;\;\;\; \psi(UCN)_{(vir)} \circ 
     (\tau_{H,-1}^{1}  \otimes \tau_{V,0}^{2})_{II} \Lra  \psi(UUN)
     \eea  
     
We get the merging the doublet AGR   with the
  quartet UCN  giving rise to the third sextet (encoding Ser).
  
\end{enumerate}

\section{ Conclusions}

The outcome of the proposed mathematical model is a pattern of
organisation in: 3 sextets, 5 quartets, 13 doublets which is
very close to the pattern of the VMC and SUC codes. In particular
it differs from the last one for the absence of the breaking of two doublets
into singlets, which may reasonably be seen as a minor effects. A more
refined analysis, see \cite{AS}, indeed gives hints for:
\begin{itemize}
\item the breaking of the doublets
\item  the choice of the Stop codons
\item the similarity of the physical chemical properties of a.a. encoded
by codons prone to misreading
\end{itemize}
 
The number and dimension of multiplets seem to be the outcome
of a {\emph strategy} addressed to keep as many as different amino acids 
with a reasonable protection against translation errors.

As final remark, in \cite{AS}, a discussion of the dependence of the 
obtained results from the choice of tensor operators mimicking the
misreading shows that there is a dependence, but the main bulk is left
unmodified.

\begin{table} 
 \footnotesize
 \centering 
 \caption{ Relative frequency (R.f.) ($10^{-3}$) of occurrence of 20 amino-acids 
(a.a.) and the number (N) of corresponding encoding codons.}
\label{tablefeq}
\begin{tabular}{|c|c|c||c|c|c||c|c|c||c|c|c|}
\hline
a.a.  &  R.f. & N & a.a.  &  R.f. & N  & a.a.  &  R.f. & N & a.a.  &  R.f. & N \\
	\hline  
	\textbf{Leu} & \textbf{91} & 6 & \textbf{Glu} & \textbf{62} & 2  &
	\textbf{Arg} & \textbf{51} & 6 & \textbf{Tyr} & \textbf{32} & 4 \\
	\textbf{Ala} & \textbf{77} & 4 & \textbf{Thr} & \textbf{59} & 4  &
	\textbf{Pro} & \textbf{51} & 4 & \textbf{Met} & \textbf{24} & 1 \\
	\textbf{Gly} & \textbf{74} & 4 & \textbf{Lys} & \textbf{59} & 2  &
    \textbf{Asn} & \textbf{43} & 2 & \textbf{His} & \textbf{23} & 2 \\
	\textbf{Ser} & \textbf{69} & 6 & \textbf{Ile} & \textbf{53} & 3 &
	\textbf{Gln} & \textbf{41} & 2 & \textbf{Cys} & \textbf{20} & 2 \\
	\textbf{Val} & \textbf{66} & 4 & \textbf{Asp} & \textbf{52} & 2 & 
	\textbf{Phe} & \textbf{40} & 2 & \textbf{Trp} & \textbf{14} & 1  \\
	 \hline  
\end{tabular}
\end{table} 

\newpage 

\begin{table}[htbp]
\caption{The vertebral mitochondrial code. The upper label denotes 
different irreducible
representations. In bold character the amino acids which are encoded 
dfferently  in the eukaryotic or standard code: UGA, AUA, AGR encoding
 respectively for Ter, Ile and Arg.  }
\label{tablerep}
\footnotesize
\begin{center}
\begin{tabular}{|cc|cc|rr|cc|cc|rr|}
\hline
codon & a.a. & $J_{H}$ & $J_{V}$ & $J_{3,H}$ & $J_{3,V}$& codon & 
a.a. & 
$J_{H}$ & $J_{V}$ & $J_{3,H}$ & $J_{3,V}$ \\
\hline
\Big. CCC & Pro & $\Half$ & $\Half$ & $\Half$ & $\Half$ & UCC & Ser & 
$\Half$ & 
$\Half$ & $\half$ & $\Half$ \\
\Big. CCU & Pro & $(\half$ & $\Half)^1$ & $\half$ & $\Half$ & UCU & 
Ser & 
$(\half$ & $\Half)^1$ & $-\half$ & $\Half$ \\
\Big. CCG & Pro & $(\Half$ & $\half)^1$ & $\Half$ & $\half$ & UCG & 
Ser & 
$(\Half$ & $\half)^1$ & $\half$ & $\half$ \\
\Big. CCA & Pro & $(\half$ & $\half)^1$ & $\half$ & $\half$ & UCA & 
Ser & 
$(\half$ & $\half)^1$ & $-\half$ & $\half$ \\[1mm]
\hline
\Big. CUC & Leu & $(\half$ & $\Half)^2$ & $\half$ & $\Half$ & UUC & 
Phe & 
$\Half$ 
& $\Half$ & $-\half$ & $\Half$ \\
\Big. CUU & Leu & $(\half$ & $\Half)^2$ & $-\half$ & $\Half$ & UUU & 
Phe & 
$\Half$ 
& $\Half$ & $-\Half$ & $\Half$ \\
\Big. CUG & Leu & $(\half$ & $\half)^3$ & $\half$ & $\half$ & UUG & 
Leu & 
$(\Half$ & $\half)^1$ & $-\half$ & $\half$ \\
\Big. CUA & Leu & $(\half$ & $\half)^3$ & $-\half$ & $\half$ & UUA & 
Leu & 
$(\Half$ & $\half)^1$ & $-\Half$ & $\half$ \\[1mm]
\hline
\Big. CGC & Arg & $(\Half$ & $\half)^2$ & $\Half$ & $\half$ & UGC & 
Cys & 
$(\Half$ & $\half)^2$ & $\half$ & $\half$ \\
\Big. CGU & Arg & $(\half$ & $\half)^2$ & $\half$ & $\half$ & UGU & 
Cys & 
$(\half$ & $\half)^2$ & $-\half$ & $\half$ \\
\Big. CGG & Arg & $(\Half$ & $\half)^2$ & $\Half$ & $-\half$ & UGG & 
Trp & 
$(\Half$ & $\half)^2$ & $\half$ & $-\half$ \\
\Big. CGA & Arg & $(\half$ & $\half)^2$ & $\half$ & $-\half$ & UGA & 
{\bf Trp} & 
$(\half$ & $\half)^2$ & $-\half$ & $-\half$ \\[1mm]
\hline
\Big. CAC & His & $(\half$ & $\half)^4$ & $\half$ & $\half$ & UAC & 
Tyr & 
$(\Half$ & $\half)^2$ & $-\half$ & $\half$ \\
\Big. CAU & His & $(\half$ & $\half)^4$ & $-\half$ & $\half$ & UAU & 
Tyr & 
$(\Half$ & $\half)^2$ & $-\Half$ & $\half$ \\
\Big. CAG & Gln & $(\half$ & $\half)^4$ & $\half$ & $-\half$ & UAG & 
Ter & 
$(\Half$ & $\half)^2$ & $-\half$ & $-\half$ \\
\Big. CAA & Gln & $(\half$ & $\half)^4$ & $-\half$ & $-\half$ & UAA & 
Ter 
& 
$(\Half$ & $\half)^2$ & $-\Half$ & $-\half$ \\[1mm]
\hline
\Big. GCC & Ala & $\Half$ & $\Half$ & $\Half$ & $\half$ & ACC & Thr & 
$\Half$ & 
$\Half$ & $\half$ & $\half$ \\
\Big. GCU & Ala & $(\half$ & $\Half)^1$ & $\half$ & $\half$ & ACU & 
Thr & 
$(\half$ & $\Half)^1$ & $-\half$ & $\half$ \\
\Big. GCG & Ala & $(\Half$ & $\half)^1$ & $\Half$ & $-\half$ & ACG & 
Thr & 
$(\Half$ & $\half)^1$ & $\half$ & $-\half$ \\
\Big. GCA & Ala & $(\half$ & $\half)^1$ & $\half$ & $-\half$ & ACA & 
Thr & 
$(\half$ & $\half)^1$ & $-\half$ & $-\half$ \\[1mm]
\hline
\Big. GUC & Val & $(\half$ & $\Half)^2$ & $\half$ & $\half$ & AUC & 
Ile & 
$\Half$ 
& $\Half$ & $-\half$ & $\half$ \\
\Big. GUU & Val & $(\half$ & $\Half)^2$ & $-\half$ & $\half$ & AUU & 
Ile & 
$\Half$ 
& $\Half$ & $-\Half$ & $\half$ \\
\Big. GUG & Val & $(\half$ & $\half)^3$ & $\half$ & $-\half$ & AUG & 
Met & 
$(\Half$ & $\half)^1$ & $-\half$ & $-\half$ \\
\Big. GUA & Val & $(\half$ & $\half)^3$ & $-\half$ & $-\half$ & 
AUA & {\bf Met} & 
$(\Half$ & $\half)^1$ & $-\Half$ & $-\half$ \\[1mm]
\hline
\Big. GGC & Gly & $\Half$ & $\Half$ & $\Half$ & $-\half$ & AGC & Ser 
& 
$\Half$ & 
$\Half$ & $\half$ & $-\half$ \\
\Big. GGU & Gly & $(\half$ & $\Half)^1$ & $\half$ & $-\half$ & AGU & 
Ser & 
$(\half$ & $\Half)^1$ & $-\half$ & $-\half$ \\
\Big. GGG & Gly & $\Half$ & $\Half$ & $\Half$ & $-\Half$ & AGG & 
{\bf Ter} & $\Half$ & 
$\Half$ & $\half$ & $-\Half$ \\
\Big. GGA & Gly & $(\half$ & $\Half)^1$ & $\half$ & $-\Half$ & 
AGA & {\bf Ter} & 
$(\half$ & $\Half)^1$ & $-\half$ & $-\Half$ \\[1mm]
\hline
\Big. GAC & Asp & $(\half$ & $\Half)^2$ & $\half$ & $-\half$ & AAC & 
Asn & 
$\Half$ 
& $\Half$ & $-\half$ & $-\half$ \\
\Big. GAU & Asp & $(\half$ & $\Half)^2$ & $-\half$ & $-\half$ & AAU & 
Asn 
& $\Half$ 
& $\Half$ & $-\Half$ & $-\half$ \\
\Big. GAG & Glu & $(\half$ & $\Half)^2$ & $\half$ & $-\Half$ & AAG & 
Lys & 
$\Half$ 
& $\Half$ & $-\half$ & $-\Half$ \\
\Big. GAA & Glu & $(\half$ & $\Half)^2$ & $-\half$ & $-\Half$ & AAA & 
Lys 
& $\Half$ 
& $\Half$ & $-\Half$ & $-\Half$ \\[1mm]
\hline
\end{tabular}
\end{center}
\end{table}

\end{document}